\documentclass[twocolumn,preprintnumbers,showpacs,amsmath,amssymb,pra]{revtex4}

\usepackage{graphicx,amsmath}
\usepackage{dcolumn}
\usepackage{bm}
\usepackage{amssymb}
\usepackage{epstopdf}
\usepackage{color}

\begin{document}

\title{Single photon superradiant decay of cyclotron resonance
in a p-type single-crystal semiconductor film with a cubic
structure }

\begin{abstract}
We study a single-photon superradiance under the conditions of
cyclotron resonance in a perfect single-crystal p-type
semiconductor film with a cubic structure. We show that the rate
of superradiant emission scales as the film area, which allows one
to specify the size of the film, at which the probability of a
single photon superradiance becomes much greater than the
probabilities of other scattering channels. The power of
superradiant emission depends only on three fundamental constants:
the electron charge $q_e$, the speed of light $c$, the electron
mass $m_e$, and on the electric to magnetic field ratio.

\end{abstract}


\date{\today }
\author{A. G. Moiseev}
\affiliation{Novosibirsk State Technical University, Novosibirsk,
Russia}

\author{Ya. S. Greenberg}\email{yakovgreenberg@yahoo.com}
\affiliation{Novosibirsk State Technical University, Novosibirsk,
Russia}


 \maketitle

\section{Introduction}\label{intr}

The use of cyclotron resonance for the study of semiconductors
began in the middle of the last century \cite{Dress53,
Dress55,Latt56,Suzuki74,Ivanov78}. Since then, the cyclotron
resonance has become a powerful tool for studying the structure of
semiconductors, allowing investigation of their band structure,
mechanisms of charge carriers scattering , the influence of
phonon-electron and phonon-hole interactio on their effective
masses, and more (see review papers \cite{Kano06,Drach12,Peet06}
and the references therein). In recent years, the development of
submicron technologies paved way for new methods of preparing
low-dimensional semiconductor structures where quantum-size
effects play a decisive role \cite{Horing13}. It makes possible
the use of cyclotron resonance for the study of the collective
effect such as Dicke superradiance, which has recently been
observed experimentally in  ultrahigh-mobility two dimensional
(2D) electron gas in GaAs\cite{Zhang14,Zhang16}, and for
electronic excitations in the InGaAs quantum well\cite{Laurent15}.

The effect of superradiation, which has been well known for a long
time (see review paper \cite{Cong16} and references therein), was
discovered by R. Dicke \cite{Dicke54}, who showed that the system
of $N$ identical two level excited atoms undergoes a spontaneous
coherent transition to the ground state. This is accompanied by
the emission of $N$ photons, the intensity which scales as $N^2$,
and the decay rate of which is $N\gamma$, where $\gamma$ is the
decay rate of an isolated atom.  As was noticed in \cite{Dicke54},
superradiant transition becomes possible if the system size $L$ is
much less than the photon wavelength $\lambda$ ($L<<\lambda$).

Another kind of superradiance (so called a single photon
superradiance) can occur when a single photon Dicke state is
formed: $N$ identical two level atoms are in a symmetrical
superposition of states with one excited atom and $N-1$ atoms in
the ground state
\cite{Scully09,Scully06,Svidz08a,Scull09,Svidz08b,Svidz09,Nef17}.
In this case, the decay rate of a single photon is also equal to
$N\gamma$. As was shown in \cite{Scully06, Svidz09}, a single
photon superradiance can occur even if system size $L$ is much
greater than the photon wavelength $\lambda$. In this case, the
photon decay rate also scales as $N$ and the photon's emission
results in a narrow radiation pattern.

Our paper is devoted to the study of a single-photon superradiance
under conditions of cyclotron resonance in a single-crystal
semiconductor film with a cubic structure. It is assumed that the
temperature is sufficiently low, so that there are no holes at the
excited Landau level ($n=1$), and the surface density of the holes
at  the lower Landau level ($n=0$) is equal to
${q_eB}/{2\pi\hbar}$. Such a density of 2D holes results in the
integer quantum Hall effect \cite{Klitzing80}, where the Hall
resistance of a semiconductor structure with 2D electron gas is
quantized and depends only on fundamental constants- the electron
charge and the Planck's constant.

In general, a superradiant transition in solids is difficult to
observe, due to inherently fast decay channels for carriers. In
semiconductors the main scattering channel for the electron and
holes is the phonon channel. The time scale of the phonon
relaxation of carriers in semiconductors, is typically of the
order of $10^{-13}s$ \cite{Yu02}.

We show in the paper that, under the conditions of cyclotron
resonance, the rate of the emission of one photon from a single
photon Dicke state is much greater than the probability of other
hole scattering mechanisms, and hence, in this case, a single
photon superradiance is the main relaxation mechanism. For
example, for the static magnetic field $B=10$ T \cite{ref} and
film size $L>0.2$ cm, the rate of a single photon superradiance in
Ge film is more than $10^{14}s^{-1}$. This value is an order of
magnitude greater than the rate of hole scattering on phonons in a
semiconductor ($10^{13}s^{-1}$) \cite{Yu02}. Therefore, under
these conditions, the emission of phonons can be neglected, and
the relaxation time is determined only by the  mechanism of a
single photon superradiance.

We also investigate the conduction and power of superradiant
emission of the two-dimensional hole gas and show that in this
case the overall universal power generated in the film depends
only on three fundamental constants $q_e, c, m_e$ and on the ratio
of intensities of the electric and magnetic fields.

The paper is organized as follows. In Section II  we describe the
cyclotron resonance spectrum of holes in a 3D single crystal of Ge
or Si in a strong homogeneous magnetic field and calculate the
rate of spontaneous  photon emission for a hole transition between
Landau levels $n = 1$ and $n = 0$. In section III we calculate the
rate of a single photon superradiance and show that the system
wave function is a symmetric superposition of single hole state
products. In Section IV we calculate a surface current. The power
of superradiant emission and its radiation patterns are found in
Section V.

\section{The cyclotron energies of the holes}

We assume the film surface is oriented in $x-y$ plane, so that
$z$- axis is directed along the [001] crystal axis. In order to
study the cyclotron resonance it is necessary to know the energy
spectrum of the holes. We take this spectrum similar to that in a
3D single crystal. As is known, the wave function of the hole is a
bispinor \cite{Latt56}. Accordingly, in 3D Si or Ge single
crystals located in a strong homogeneous magnetic field $B$
applied along the [001] axis, there are four energy levels for the
first two Landau levels n = 0,1. In the framework of perturbation
theory these energies were calculated in \cite{Moiseev15} up to
the second order of magnitude under conditions ${\hbar
^2}k_z^2/2{m_e} <  < \hbar {\omega _c};\quad \mu  = 0.5\left(
{{\gamma _3} - {\gamma _2}} \right) < 1$, where ${\omega _c} =
{q_e}B/{m_e}$ is the cyclotron frequency, to be

\begin{equation}\label{en}
E_{\alpha ,n}^{} = E_{\alpha ,n}^{\left( 0 \right)} + E_{\alpha
,n}^{\left( 1 \right)} + E_{\alpha ,n}^{\left( 2 \right)}
\end{equation}
where the first subscript numbers the bispinor ($\alpha=1,2$), and
the second subscript numbers the Landau levels ($n=0,1$).

For the subsequent study it is important that the energy spectrum
 of the holes in Ge and Si (\ref{en}) is not equidistant relative
to the quantum number $n$ \cite{Latt56} and the energy
$E_{\alpha,n}$ is independent on the quantum number $k_x$
\cite{Moiseev15}. Expression (\ref{en}) can be used for the
calculation of hole energies in a film under the condition \cite{
Moiseev15}:
\begin{equation}\label{cond}
\frac{{{\pi ^2}{\hbar ^2}}}{{2{m_{\alpha ,n}}}}\frac{{{{\left(
{n'} \right)}^2}}}{{{d^2}}} <  < \hbar {\omega _c}
\end{equation}
where $m_{\alpha,n}$ is the effective mass of a hole in 3D single
crystal \cite{ Moiseev15}, $d$ is the film thickness, $n'$ is the
number of de Broglie half waves across the film.

Condition (\ref{cond}) holds to a good accuracy for a magnetic
field $B=10$ T, film thickness $d=2.0\times 10^3\AA$, and $n'=1$.
It allows one to take zero approximation in (\ref{en}),
$E_{\alpha,n}^{(0)}$ for the calculation of the energy spectrum
\cite{ Moiseev15}:

\begin{subequations}
\begin{equation}\label{en1}
E_{1,0}^{(0)} = \frac{1}{2}\hbar {\omega _c}\left( {{\gamma _2} -
{\gamma _1} + k} \right)
\end{equation}
\begin{equation}\label{en2}
E_{1,1}^{(0)} = \frac{1}{2}\hbar {\omega _c}\left( {3({\gamma _2}
- {\gamma _1}) + k} \right)
\end{equation}
\end{subequations}

\begin{subequations}
\begin{equation}\label{en3}
E_{2,0}^{(0)} =  - \frac{1}{2}\hbar {\omega _c}\left( {{\gamma _2}
+ {\gamma _1} - 3k} \right)
\end{equation}
\begin{equation}\label{en4}
E_{2,1}^{(0)} =  - \frac{3}{2}\hbar {\omega _c}\left( {{\gamma _2}
+ {\gamma _1} - k} \right)
\end{equation}
\end{subequations}
where $\gamma_1,\gamma_2, \gamma_3$ are the Lattinger parameters .
In Ge $\gamma_1=13.2, \gamma_2=4.4, \gamma_3=5.4 $\cite{Latt56},
$k=-3.41$ \cite{Bir74}; in Si $\gamma_1=4.22, \gamma_2=0.5,
\gamma_3=1.38$ \cite{Bir74}.

The eigenvectors for energies
(\ref{en1}),(\ref{en2}),(\ref{en3}),(\ref{en4}) take on the form:

\begin{equation}\label{egn}
\left\langle {\psi _{\alpha ,n}^{(0)}} \right| = \left( {0,(2 -
\alpha )u_n^ * ,0,(\alpha  - 1)u_n^ * } \right)
\end{equation}
where $u_n$ is the spatial part of the wavefunction:
\begin{equation}\label{space}
{u_n}({k_x},x,y) = {C_n}\sqrt {\frac{1}{{{L_x}d}}} {e^{i{k_x}x}}
{e^{ - \frac{{\xi _{}^2}}{2}}}{H_n}({\xi _{}})
\end{equation}
where $\xi  = \sqrt {\frac{{{m_e}{\omega _c}}}{\hbar }} \left( {y
- R_e^2{k_x}} \right)$, ${C_n} = \frac{1}{{\sqrt {{2^n}n!\sqrt \pi
{R_e}} }}$, ${R_e} = \sqrt {\frac{{\hbar}}{{{q_e}B}}}$ is a
cyclotron radius, $H_n(\xi)$ are the Hermite polynomials, $L_x$ is
the film length along the $x$ axis, and ${k_x} = \frac{{2\pi
}}{{{L_x}}}{n_x}$, ${n_x} = 0, \pm 1, \pm 2,...$.

The resonance transition is possible only between different Landau
levels $n$ which belongs to the same bispinor. From expressions
(\ref{en1}),(\ref{en2}),(\ref{en3}),(\ref{en4}) we obtain the
frequencies of the corresponding transitions:
\begin{equation}\label{freq}
\hbar {\omega _\alpha } = E_{\alpha ,0}^{\left( 0 \right)} -
E_{\alpha ,1}^{\left( 0 \right)} = \hbar {\omega _c}{C_\alpha
};\quad \alpha  = 1,2
\end{equation}
where ${C_\alpha } = \left( {{\gamma _1} + {{( - 1)}^\alpha
}{\gamma _2}} \right)$.

In general, our approach is valid when $R_e \gg a_0$, where $a_0$
is the lattice constant (for Ge $a_0=5.6\AA$). From $R_e = a_0$ we
estimate the maximal value of the magnetic field to be
$B_0=2.1\times 10^3$ T. Therefore, our scheme for the calculation
of the holes' spectrum is justified for $B\ll B_0$. On the other
hand, the expressions (\ref{en1}), (\ref{en2}), (\ref{en3}), and
(\ref{en4}) provide a good approximation if $|E^{(0)}_{\alpha
,n}/\Delta|< 1$, where $\Delta$ is the spin-orbit splitting (for
Ge $\Delta=0.29$ eV). The calculations for Ge show that for
magnetic fields $B=(1\div 10)$ T the ratio $|E^{(0)}_{\alpha
,n}/\Delta|$ does not exceed 0.12.

\subsection{The rate of spontaneous photon emission under hole
transition between $\textbf{n=1}$ and $\textbf{n=0}$ Landau
levels.}

In the dipole approximation, the rate $\Gamma_\alpha$ for the hole
transition between states $\left| {\psi _{\alpha ,1}^{\left( 0
\right)}} \right\rangle $ and $\left| {\psi _{\alpha ,0}^{\left( 0
\right)}} \right\rangle$ with the emission of  a photon can be
obtained from the conventional expression:

\begin{equation}\label{Gamma}
{\Gamma _\alpha } = \frac{{\omega _\alpha
^3}}{{3\pi\varepsilon_0\hbar {c^3}}}{\left| {\left\langle {\psi
_{\alpha ,1}^{\left( 0 \right)}} \right|{q_e}\hat y\left| {\psi
_{\alpha ,0}^{\left( 0 \right)}} \right\rangle } \right|^2}
\end{equation}
where $\left\langle {\psi _{\alpha ,1}^{\left( 0 \right)}}
\right|{q_e}\hat y\left| {\psi _{\alpha ,0}^{\left( 0 \right)}}
\right\rangle  = {q_e}{R_e}\frac{1}{{\sqrt 2 }}{\delta
_{{k_x},{{k'}_x}}}$, $\varepsilon_0$ is the electric constant.

Finally for $\Gamma_{\alpha}$ we obtain:
\begin{equation}\label{Gamma1}
{\Gamma _\alpha } = \frac{C_\alpha ^3}{{6\pi\varepsilon_0{{\left(
{2\pi } \right)}^3}}}\frac{{q_e^2}}{{{R_e}\hbar }}{\left(
{\frac{{{\lambda _C}}}{{{R_e}}}} \right)^3}
\end{equation}
where ${\lambda _C} = 2\pi \hbar /{m_e}c$ is the electron Compton
wavelength.

The lifetime of the state $\left| {\psi _{\alpha ,1}^{\left( 0
\right)}} \right\rangle$ is given by the quantity
$\tau_\alpha=1/\Gamma_\alpha$. For the magnetic field strength
$B=10$ T we obtain from (\ref{Gamma1}) the corresponding lifetimes
$\tau_1=7.6\times 10^{-5}$ s, $\tau_2=9.5\times 10^{-6}$ s. These
values are much greater than the lifetime of state $\left| {\psi
_{\alpha ,1}^{\left( 0 \right)}} \right\rangle$ against a phonon
emission which is of the order of $10^{-13}$ s in
semiconductors\cite{Yu02}. It would seem that under these
conditions the photon decay channel is impossible. However, we
will show in the next sections that due to the mechanism of a
single photon superradiance, the decay channel of the state
$\left| {\psi _{\alpha ,1}^{\left( 0 \right)}} \right\rangle$
against the photon emission becomes the dominating process.

\section{Single photon superradiance}\label{spr}

In order to estimate the rate of single photon superradiance we
use the method of non-Hermitian effective Hamiltonian
\cite{Auerrbach11}, which has been applied to the study of
microwave scattering on a chain of two level atoms
\cite{Greenberg15}. We consider a one dimensional chain of $N$
noninteracting holes aligned along the $y$- axis with the incident
photon directed along the $z$-axis. As a basis set of state
vectors we take the states where one hole is in the excited state
$|e\rangle$ and other $N-1$ holes are in the ground state
$|g\rangle$. Therefore, we have $N$ vectors
$|n\rangle=|g_1,g_2,.....g_{n-1},e_n,g_{n+1},....g_{N-1},g_N\rangle$.
The spontaneous emission of the excited hole results in a
continuum of states
$|k\rangle=|g_1,g_2,.......g_{N-1},g_N,k\rangle$, where all holes
are in the ground state and there is one photon in the system.
This process can be described by non Hermitian Hamiltonian:
\begin{equation}\label{nonH}
    H=H_0-iW
\end{equation}
where $H_0$ is hamiltonian of holes, and operator $W$ describes
the interaction of the holes with the photon field.

The matrix elements of (\ref{nonH}) in the {$|n\rangle$}
representation are:
\begin{equation}\label{me}
\left\langle m \right|H\left| n \right\rangle  = \hbar {\omega
_\alpha }{\delta _{m,n}} - i\left\langle m \right|W\left| n
\right\rangle ; (1 \le m,n \le N).
\end{equation}
If the distance between holes along the direction of the photon
scattering ($z$-axis) is much less than the photon wavelength, the
matrix element on the right hand side of (\ref{me}) takes the form
\cite{Greenberg15}:

\begin{equation}\label{W}
\left\langle m \right|W\left| n \right\rangle  = \hbar \sqrt
{\Gamma _\alpha ^{(m)}\Gamma _\alpha ^{(n)}}
\end{equation}
where $\Gamma _\alpha ^{(n)}$ is the rate of  spontaneous photon
emission from a state where the $n$-th hole is excited.

Due to the planar geometry of the film, the z coordinates of all
holes in the chain are the same- they are in an identical
arrangement relative to a wavefront. Therefore, we assume that the
rate of spontaneous emission of holes is the same: $\left\langle m
\right|W\left| n \right\rangle = \hbar \Gamma _\alpha$. Thus, we
get a non-Hermitian $N\times N$ matrix, where the main diagonal
elements are $\hbar {\omega _\alpha } - i\hbar {\Gamma _\alpha }$,
and all off-diagonal elements are equal to $ - i\hbar {\Gamma
_\alpha }$:
\begin{equation}\label{Matrix}
\begin{array}{l}
\frac{1}{\hbar }\left\langle m \right|H\left| n \right\rangle
\\ =\left( {\begin{array}{*{20}{c}}
{{\omega _\alpha } - i{\Gamma _\alpha }}&{ - i{\Gamma _\alpha }}&{ - i{\Gamma _\alpha }}& \ldots &{ - i{\Gamma _\alpha }}\\
{ - i{\Gamma _\alpha }}&{{\omega _\alpha } - i{\Gamma _\alpha }}& \ldots & \ldots &{ - i{\Gamma _\alpha }}\\
{ - i{\Gamma _\alpha }}&{ - i{\Gamma _\alpha }}&{{\omega _\alpha } - i{\Gamma _\alpha }}& \ldots &{ - i{\Gamma _\alpha }}\\
 \vdots & \vdots & \vdots & \ddots & \vdots \\
{ - i{\Gamma _\alpha }}&{ - i{\Gamma _\alpha }}& \ldots & \ldots
&{{\omega _\alpha } - i{\Gamma _\alpha }}
\end{array}} \right)
\end{array}
\end{equation}
The incident photon, when absorbed by the film, can excite any
hole. As we do not know which of $N$ holes is excited, the wave
function of the holes should be expressed as a superposition of
the state vectors $|n\rangle$:
\begin{equation}\label{wf}
\Psi  = \sum\limits_{n = 1}^N {{c_n}\left| n \right\rangle }
\end{equation}
It is not difficult to show that the solution of the Schr\"odinger
equation $H\Psi=E\Psi$, with $H$ and $\Psi$ from (\ref{Matrix})
and (\ref{wf}) respectively, has the following properties: 1.
There is a single state with energy ${E_S} = \hbar {\omega _\alpha
} - i\hbar N{\Gamma _\alpha }$ whose wavefunction is a symmetric
coherent superposition of the state vectors $|n\rangle$, where all
quantities $c_n$ are the same:
\begin{equation}\label{Sym}
{|\Psi _S\rangle} = \frac{1}{{\sqrt N }}\sum\limits_{n = 1}^N
{\left| n \right\rangle }
\end{equation}
2. There are $N-1$ degenerate states with energy $E = \hbar
{\omega _\alpha }$, where all coefficients $c_n$ in (\ref{wf})
satisfy the condition $\sum\limits_{n = 1}^N {{c_n}}  = 0$. These
states are dark, non decaying states since their widths are equal
to zero.

The collective state $|\Psi_S\rangle$ (\ref{Sym}), which we call a
single photon Dicke state, can be formed by a single photon, which
propagates normal to the film surface and interacts in-phase with
every hole in the plane of the film \cite{Scully06}.

 Therefore, under this conditions, state (\ref{Sym}) decays with a rate
 $N\Gamma_\alpha$, so that the rate of
the spontaneous emission of a single hole $\Gamma_{\alpha}$
(\ref{Gamma1}) should be substituted with the quantity
$\overline{\Gamma}_{\alpha}$:
\begin{equation}\label{Gamma}
{\bar \Gamma _\alpha } = N{\Gamma _\alpha } = \frac{{2\pi
}}{3}\frac{{q_e^2}}{{{\varepsilon _0}\hbar {\lambda _\alpha
}}}{\left( {\frac{L}{{{\lambda _\alpha }}}} \right)^2}
\end{equation}
which is the linewidth of a single photon superradiant emission.
In Eq.\ref{Gamma} the quantity $\lambda_{\alpha}$ ($\alpha=1,2$)
is the wavelength of emitted photon:

\begin{equation}\label{lw}
{\lambda _\alpha } = \frac{{2\pi c{m_e}}}{{{C_\alpha }{q_e}B}},
\end{equation}
and $N$ is the number of holes which take part in the formation of
the single photon Dicke state (\ref{Sym}):

\begin{equation}\label{nh}
{N} = \frac{{{q_e}B}}{{2\pi \hbar}}{L^2} = \frac{1}{{2\pi
}}{\left( {\frac{L}{{{R_e}}}} \right)^2}
\end{equation}
where $L=L_x=L_y$.

From the considerations given above we obtain the following
estimations. For a magnetic field $B=10$ T we estimate transition
frequencies (\ref{freq}) $\omega_1=1.6\times 10^{13}$rad/s,
$\omega_2=3.1\times 10^{13}$rad/s, with  corresponding wavelengths
$\lambda_1=0.012$ cm and $\lambda_2=0.006$ cm. In the range $0.2$
cm$\leq L\leq 0.4$ cm the expression (\ref{nh}) gives $9.7\times
10^{9}\leq N\leq 3.9\times 10^{10}$. Then, from expression
(\ref{Gamma}), it follows that the rate of spontaneous hole
emission from a Ge film is more than $10^{14}s^{-1}$. Since the
rate of the phonon scattering in semiconductors is of the order of
$10^{13}s^{-1}$ (see, for example,\cite{Yu02}), we may neglect all
scattering mechanisms except for a single photon superradiance,
which becomes, under these conditions, the main relaxation
mechanism of excited holes.

\section{The surface current}\label{surfcurr}

First we define the ground state $|G\rangle$ of the ensemble of
the holes: $\left| G \right\rangle = \left|
{{g_1},.{g_2},.....{g_n},.......{g_{N - 1}},{g_N}} \right\rangle $
with the energy $\varepsilon_G$. Next we take the external time
dependent electric field, which is directed normal to the time
independent homogeneous strong magnetic field:
\begin{equation}\label{drive}
\hat V\left( {y,t} \right) = \left\{ {\begin{array}{*{20}{c}}
{0,_{}^{}t < 0}\\
{\hat V\cos \left( {\omega t} \right),_{}^{}t > 0}
\end{array}} \right.
\end{equation}
where $\hat V =  - y{q_e}{E_y}$. It is not difficult to show that
this driving field gives rise to transitions only between the
states $|G\rangle$ and $|\Psi_S\rangle$. The matrix element of the
dipole operator between these states is: $\left\langle {{\Psi _S}}
\right|{q_e}\hat y\left| G \right\rangle = {q_e}{R_e}\sqrt {N/2}
$, while the transition amplitudes between $|G\rangle$ and dark
states are zero.

As the energy spectrum of holes in Ge and Si is not equidistant
relative to the quantum number $n$ \cite{Latt56}, the evolution of
the hole state vector $|\Psi(t)\rangle$, which accounts for the
near resonant transitions at $\hbar\omega_{\alpha}>>k_BT$, between
states $|G\rangle$ and $|\Psi_S\rangle$ is as follows:

\begin{equation}\label{evol}
\left| {\Psi (t)} \right\rangle  = \left| G \right\rangle a\left(
t \right){e^{ - i\frac{{{\varepsilon _G}}}{\hbar }t}} + \left|
{{\Psi _S}} \right\rangle b\left( t \right){e^{ - i\left(
{\frac{{{\varepsilon _G}}}{\hbar } + {\omega _\alpha } - i{{\bar
\Gamma }_\alpha }} \right)t}}
\end{equation}
where the amplitudes $a(t)$ and $b(t)$ satisfy initial conditions
$a(0)=1$, $b(0)=0$.

These amplitudes can be found near resonance
$\omega\approx\omega_{\alpha}$, in the frame of conventional time
dependent perturbation theory: $a(t)=1$,

\begin{equation}\label{b}
b\left( t \right) = {E_y}\frac{{{q_e}{R_e}\sqrt N }}{{2\sqrt 2
\hbar }}\left( {\frac{{{e^{i\left( {{\omega _\alpha } - i{{\bar
\Gamma }_\alpha } - \omega } \right)t}} - 1}}{{\left( {{\omega
_\alpha } - i{{\bar \Gamma }_\alpha } - \omega } \right)}}}
\right)
\end{equation}
At resonance, $\omega=\omega_{\alpha}$, the condition $|b(t)|\ll
1$ sets an upper bound on the amplitude of the external electric
field $E_y$: $\left({E_y} < < \frac{{2\sqrt{2} \hbar
{\overline{\Gamma} _\alpha } }}{{{q_e}{R_e}}\sqrt{N}}\equiv
E_{\alpha }^{\max }\right)$.

From equation (\ref{evol}) we calculate a time dependent steady
state part of the hole dipole moment $\mathop {\lim }\limits_{t
\to \infty } \left\langle {\Psi (t)} \right|{q_e}\hat y\left|
{\Psi (t)} \right\rangle  \equiv \left\langle {{q_e} y(t)}
\right\rangle $, which causes the transitions between states
$|G\rangle$ and $|\Psi_S\rangle$

\begin{equation}\label{steady}
\langle q_e y(t)\rangle= {E_y}\frac{N{{{\left( {{q_e}{R_e}}
\right)}^2}}}{{2\hbar }}\left( {\frac{{\left( {{\omega _\alpha } -
\omega } \right)\cos \omega t + {\bar\Gamma _\alpha }\sin \omega
t}}{{{{\left( {{\omega _\alpha } - \omega } \right)}^2} +
\bar\Gamma _\alpha ^2}}} \right)
\end{equation}

From (\ref{steady}) we estimate the rate of change of the average
dipole moment of the holes at resonant frequency
$\omega=\omega_{\alpha}$:

\begin{equation}\label{rate}
\left\langle {{q_e}\dot y(t)} \right\rangle_r  =
{E_y}\frac{{N{{\left( {{q_e}{R_e}} \right)}^2}{\omega _{_\alpha
}}}}{{2\hbar {{\bar{\Gamma} }_\alpha }}}\cos \left( {{\omega
_\alpha }t} \right)
\end{equation}

From (\ref{rate}) we introduce the average velocity of a hole
$\langle v\rangle$: $\langle q_e \dot y(t)\rangle=Nq_e\langle
v\rangle $ and define a surface current density

\begin{equation}\label{scurr}
    J_y(\omega=\omega_{\alpha},t)=\frac{Nq_e\langle
v\rangle}{L^2}
\end{equation}

And finally, from (\ref{scurr}) we can estimate the current in the
film and the conductivity of an ideal 2D system. Synchronous
steady state motion of the holes allows us to find the
conductivity of a 2D system when the number of holes, which take
part in the formation of the single photon Dicke state
(\ref{Sym}), is equal to $N$ (\ref{nh}).

\section{The angular distribution of superradiant emission}

The total power, which is supplied to a film, gives rise to the
transitions between the states $|G\rangle$ and $|\Psi_S\rangle$,
\begin{equation}\label{Power}
{P_\alpha } = \frac{1}{2}{\sigma _\alpha }E_y^2{L^2}
\end{equation}
We assume there are no dissipative losses, so that all this power
is radiated into a free space.

In (\ref{Power}) a quantity $\sigma_{\alpha}$ is the conductivity
at the frequency $\omega=\omega_{\alpha}$, which is obtained from
(\ref{rate}) and (\ref{scurr}):
\begin{equation}\label{Conduct}
{\sigma _\alpha } = \frac{1}{{4\pi }}\frac{{q_e^2{\omega _\alpha
}}}{{\hbar N{\Gamma _\alpha }}} = \frac{3}{{4\pi }}\sqrt
{\frac{{{\varepsilon _0}}}{{{\mu _0}}}} {\left( {\frac{{{\lambda
_\alpha }}}{L}} \right)^2}
\end{equation}

Hence, we may express $P_{\alpha}$ in the following form:
\begin{equation}\label{Power1}
{P_\alpha } = \frac{3}{{8\pi }}\sqrt {\frac{{{\varepsilon
_0}}}{{{\mu _0}}}} \lambda _\alpha ^2E_y^2
\end{equation}
where, as we noted before, the amplitude of electric field
satisfies the condition $E_y\ll E_{\alpha}^{max}$.

In Ge with $B=10 $ T and $L=0.2 $ cm, the upper limit of the
electric field intensity  $E_{2}^{maz}=2.4\times 10^3 $ V/m, and
from $E_y=0.2E_{2}^{max}$ we obtain the emission power
$P_{2}=2.6\times 10^{-7}$ J/s.

It is seen from (\ref{Power1}) that the quantity
$C^2_{\alpha}P_{\alpha}$, which we call the universal emission
power, depends neither on the film dimension $L$ nor on the
material properties. It depends only on the fundamental constants
$q_e, c, m_e$ and on the electrical to magnetic field ratio
$E_y/B$.

From experimental point of view, it is important to know the
angular distribution of a radiation field. An exact form of
radiation pattern is given by the real part of the time-average
power density
$\langle\textbf{S}\rangle=\frac{1}{2\mu_0}[\textbf{E}\times
\textbf{B}^*]$, where $\textbf{E}$ and $\textbf{B}$ refer to the
peak amplitudes of the oscillating quantities,
$\textbf{E}(t)=\textbf{E}e^{i\omega t},
\textbf{B}(t)=\textbf{B}e^{i\omega t}$. In what follows, we
calculate the radiation pattern of spontaneous emission in a
far-field region ($r\gg\lambda, r\lambda\geq L^2$), where in a
single electromagnetic plane wave a vector $\textbf{E}$ is normal
to a vector $\textbf{B}$, and $E=cB$. Hence, in this region the
time-average vector power density $\langle\textbf{S}\rangle$ is
simply a real number:
$\langle\textbf{S}\rangle=\frac{c}{2\mu_0}|\textbf{B}|^2$.

The magnetic field in a far-field region is given by the
expression (see the expression (\ref{BB2}) in the Appendix):

\begin{equation}\label{B}
\textbf{B}\left( {\bf{r}} \right) =  - i\frac{{{\mu _0}}}{{4\pi
}}\left[ {{{\bf{k}}} \times {\bf{J}}\left( {{{\bf{k}}}} \right)}
\right]\frac{{{e^{i{k}r}}}}{r}.
\end{equation}
where $r$ is a distance from a source of the field, $\textbf{k}$
is the wave vector ($k=\omega/c$), which is directed along
$\textbf{r}$ in a far-field region, $\textbf{J}(\textbf{k})$ is a
spectral component of a source current $\textbf{J}(\textbf{r})$:

\begin{equation}\label{J}
{\bf{J}}\left( {\bf{k}} \right) = \int\limits_V^{} {}
{\bf{J}}\left( {\bf{r}} \right){e^{ - i\left( {{\bf{k}} \cdot
{\bf{r}}} \right)}}d{\bf{r}}
\end{equation}

Therefore, for $\langle S\rangle$ we obtain:
\begin{equation}\label{Poy}
   \langle
   S\rangle=\frac{1}{2r^2}\sqrt{\frac{\mu_0}{\varepsilon_0}}\left(\textbf{k}\times
   \textbf{J}(\textbf{k})\right)^2
\end{equation}

In our case, the current in a square $L\times L$ film can be
written as:
\begin{equation}\label{Jr}
{\bf{J}}\left( {\bf{r}} \right) = \left\{ {\begin{array}{*{20}{c}}
{0._{}^{}\quad\quad\quad\quad\left| x \right|,\left| y \right| > \frac{L}{2}}\\
{{{\rm{\textbf{e}}}_y}{J_y}\delta \left( z
\right)._{}^{}\quad\left| x \right|,\left| y \right| \le
\frac{L}{2}}
\end{array}} \right.
\end{equation}
where $J_y$ is given in (\ref{scurr}). In (\ref{Jr}) the origin of
coordinates is taken in the geometrical center of a film where the
$z$-axis is normal to the film plane. From (\ref{J}) we find the
spectral component $\textbf{J}(\textbf{k})$:

\begin{equation}\label{Jk}
{\bf{J}}\left( {\bf{k}} \right) =
{\textbf{e}_y}{J_y}{L^2}\frac{{\sin \left( {\frac{{{k_x}L}}{2}}
\right)}}{{\frac{{{k_x}L}}{2}}}\frac{{\sin \left(
{\frac{{{k_y}L}}{2}} \right)}}{{\frac{{{k_y}L}}{2}}}
\end{equation}
where
\begin{equation}\label{k}
{\bf{k}} = {\textbf{e}_x}{k_{\alpha}}\sin \left( \theta
\right)\cos \left( \varphi  \right) +
{\textbf{e}_y}{k_{\alpha}}\sin \left( \theta \right)\sin \left(
\varphi  \right) + {\textbf{e}_z}{k_{\alpha}}\cos \left( \theta
\right),
\end{equation}
$0\leq\theta\leq\pi, 0\leq\varphi\leq2\pi$,
$k_{\alpha}=2\pi/\lambda_{\alpha}$, and
$\textbf{e}_x,\textbf{e}_y,\textbf{e}_z$ are unit vectors in the
direction of the x-axis, y-axis, and $z$-axis respectively.

A substitution of (\ref{Jk}) in (\ref{Poy}) yields the radiated
power density:

\begin{equation}\label{POWD}
\left\langle {S({\rm{r}})} \right\rangle  = \sqrt {\frac{{{\mu
_0}}}{{{\varepsilon _0}}}} \frac{{{J_y}^2{L^4}}}{{4{\lambda
^2}{r^2}}}f(\theta ,\varphi ), \quad[W/m^2]
\end{equation}
where
\begin{equation}\label{NPOWD}
f(\theta ,\varphi ) = \left( {{{\cos }^2}\theta  + {{\sin
}^2}\theta {{\cos }^2}\varphi } \right){\left( {\frac{{\sin \left(
{\frac{{{k_x}L}}{2}} \right)}}{{\frac{{{k_x}L}}{2}}}\frac{{\sin
\left( {\frac{{{k_y}L}}{2}} \right)}}{{\frac{{{k_y}L}}{2}}}}
\right)^2}
\end{equation}
is the normalized power density which defines the angular
distribution of a supperradiant emission. Spherical angles
$\theta$ and $\varphi$ in (\ref{k}) and (\ref{NPOWD}) coincide
with those of vector $\textbf{r}$ since in a far-field region
vector $\textbf{k}$ is directed along $\textbf{r}$. We note that,
except for the first factor in the right hand side of
(\ref{NPOWD}), the expression for $f(\theta,\varphi)$ is similar
to that of Fraunhofer diffraction on a square aperture.

\begin{figure}
  \includegraphics[width=8 cm]{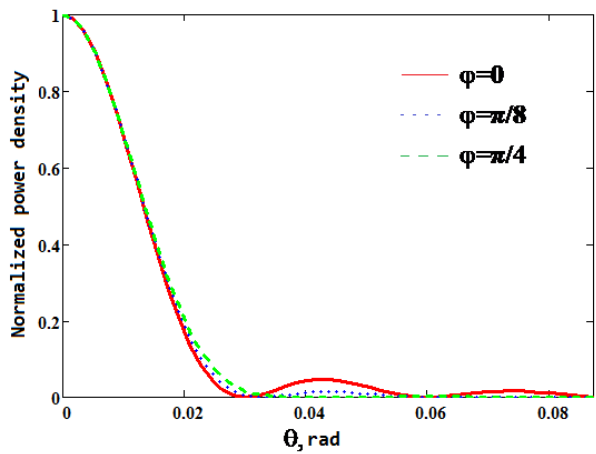}\\
  \caption{Normalized radiated power density $f(\theta,\varphi)$
  vs $\theta$ for fixed $\varphi$. $\lambda_2=0.006$ cm, $L=0.2$ cm.}\label{1D}
\end{figure}

\begin{figure}
  \includegraphics[width=8 cm]{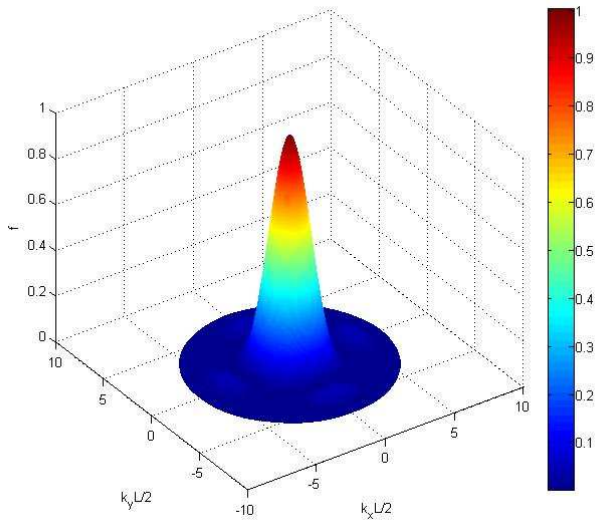}\\
  \caption{3D surface pattern of normalized radiation power density. }\label{3D}
\end{figure}

\begin{figure}
  \includegraphics[width=9 cm]{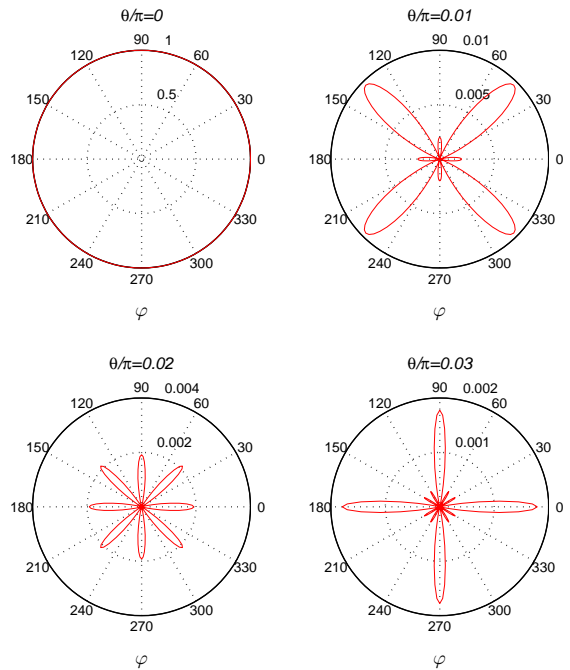}\\
  \caption{Polar patterns of normalized radiation power density.}\label{2D}
\end{figure}

In order to visualize the angle dependence of emission power
density we draw the function $f(\theta,\varphi)$ (\ref{NPOWD}) in
three different coordinates. The plots are performed for
$\lambda_2=0.006$ cm, $L=0.2$ cm, so the far-field region
corresponds to $r\geq L^2/\lambda_2\approx 6.6$ cm. In
Fig.\ref{1D} we show the dependence of normalized power density
$f(\theta,\varphi)$ on $\theta$ for several fixed polar angles
$\varphi$. A 3D plot of the normalized radiation density emitted
in the upper half space is shown in Fig.\ref{3D}. It is evident
from these plots that for our parameters most of the power is
radiated within a narrow region near a $z$-axis, which corresponds
to the solid angle
$\delta\Omega\approx\pi(\lambda_2/L)^2=2.82\times 10^{-3}$ sr. The
main and minor lobes can be seen in polar patterns of radiation
power density as shown in Fig.\ref{2D}.

\section{Discussion}

In the paper we study a single-photon superradiance under the
conditions of cyclotron resonance in a perfect single-crystal
p-type semiconductor film with a cubic structure. We assume the
film is at a sufficiently low temperature, so that we are able to
take the initial hole density at the Landau level $n=0$ to be
$q_eB/2\pi\hbar$, with no holes at the excited Landau level $n=1$.

We show that the rate of superradiant emission, which results from
the transition between the collective states $|G\rangle$ and
$|\Psi_S\rangle$, scales as the film area, which allows one to
specify the size of the film, at which the probability of a single
photon superradiance becomes much greater than the probabilities
of other scattering channels. For Ge in a static magnetic field of
the order of $10$ T  and film dimension $L>0.2$ cm, the rate of a
single photon superradiance due to a hole transition is more than
$10^{14}s^{-1}$. This value is an order of magnitude higher than
the rate of the phonon emission by a hole. Therefore, we may
neglect all scattering mechanisms except for a single photon
superradiance, which becomes, under these conditions, the main
relaxation mechanism of excited holes.

We show that the universal power of superradiant emission depends
only on the fundamental constants $q_e, c, m_e$ and on the
electric to magnetic field ratio $E_y/B$.

We calculate the angular distribution of superradiant emission and
show that for our parameters most of the power is radiated within
a narrow region near a $z$-axis, which corresponds to the solid
angle $\delta\Omega\approx\pi(\lambda_2/L)^2=2.82\times 10^{-3}$
sr.

In conclusion we would like to mention several issues which may be
important in the experimental realization of this effect.

A necessary condition for the formation of the single photon Dicke
state $|\Psi_S\rangle$ (\ref{Sym}) is the existence of a single
driving photon, which propagates normal to the film surface
\cite{Scully06}. In principle, it could be arranged if the film
under study is embedded in a resonant cavity whose fundamental
frequency is close to the transition frequency between Landau
levels $n = 1$ and $n = 0$.

We showed in the paper that in order to obtain a large decay rate
$N\Gamma_{\alpha}$, which overcomes other scattering channels, the
film size $L$ should be much greater than the photon wavelength
$\lambda$. For large samples it leads to a reduction of the decay
rate by a factor $(\lambda/L)^2$ \cite{Scull09}, due to
characteristic phase factors $e^{i\vec{k}\vec{r}_j}$, where
$\vec{k}$ is the wave vector of incident photon, $\vec{r}_j$ is
the hole position in the crystal volume. For a thin crystal film,
which we consider here, the majority of the emitters are located
near the film surface. In the case of the incident photon
propagating normal to the film surface, all surface emitters
experience nearly the same phase shift, so that in our case we may
neglect the geometrical reduction of the decay rate.

As was shown above, for the formation of the the quasi stationary
state $|\Psi_S\rangle$, with a large decay rate
$N\Gamma_{\alpha}$, the transition frequencies $\omega_{\alpha}$
and decay rates $\Gamma_{\alpha}$ for all emitters should be the
same. It means that  two-level systems (\ref{space})
$u_0(k_x,x,y), u_1(k_x,x,y)$, spaced by different $k_x$ along the
$y$- axis, need to be identical.  To ensure this condition the
film under study should be as ideal as possible. There cannot be
local defects in the film located close to the maxima of the wave
functions $u_0(k_x,x,y), u_1(k_x,x,y)$, whose positions are
determined by the magnitude of $k_x$.

We believe that the results obtained in our study will help to
open a new window for developing novel light sources based on
superradiance emission.

\begin{acknowledgements}
A. G. M. thanks M. V. Entin for many fruitful discussions. The
authors acknowledge financial support from Ministry of Education
and Science of the Russian Federation under Project No. 419
3.4571.2017/6.7.
\end{acknowledgements}

\appendix*

\section{The calculation of magnetic field in a far-field region}\label{A1}
The magnetic field generated by a source current density
$\textbf{J(\textbf{r}')}$ in an arbitrary point $\textbf{r}$ of
space can be found from Maxwell equations in the following form
\cite{Stratton}:

\begin{equation}\label{B1}
\textbf{B}({\bf{r}}) =  - {\mu _0}\int\limits_V^{} {} \left[
{{\nabla _r}G\left( {{\bf{r}} - {\bf{r'}}} \right) \times
{\bf{J}}\left( {{\bf{r'}}} \right)} \right]d{\bf{r'}}
\end{equation}
where the integration in (\ref{B1}) is over the distribution of a
source current density $\textbf{J(\textbf{r}')}$. The quantity
$G(\textbf{r}-\textbf{r}')$ is the free-space Green's function of
the scalar Helmholtz equation:

\begin{equation}\label{Green}
G\left( {{\bf{r}} - {\bf{r'}}} \right) = \frac{{{e^{ik\left|
{{\bf{r}} - {\bf{r'}}} \right|}}}}{{4\pi \left| {{\bf{r}} -
{\bf{r'}}} \right|}}
\end{equation}
where $k$ is the plane wave wavevector, $k=\omega/c=2\pi/\lambda$.

Below we use a spectral representation of Green's function
(\ref{Green})
\begin{equation}\label{Green1}
G\left( {{\bf{r}} - {\bf{r'}}} \right) = \int
{\frac{{d{\bf{k'}}}}{{{{(2\pi )}^3}}}\frac{{{e^{i{\bf{k'}} \cdot
\left( {{\bf{r}} - {\bf{r'}}} \right)}}}}{{{{k'}^2} - {k^2} -
i\varepsilon }}}
\end{equation}
where a small imaginary quantity $\varepsilon$ in the denominator
of (\ref{Green1}) ensures the outgoing scattering wave solution of
Helmholtz equation.

Substitution of (\ref{Green1}) into (\ref{B1}) yields the result:
\begin{equation}\label{BB}
\textbf{B}\left( {\bf{r}} \right) =  - i\frac{\mu_0}{{8{\pi
^3}}}\int\limits_V^{} {} \int\limits_{{\bf{k'}}}^{} {} \left[
{{\bf{k'}} \times {\bf{J}}\left( {{\bf{r'}}} \right)}
\right]\frac{{{e^{i{\bf{k'}} \cdot \left( {{\bf{r}} - {\bf{r'}}}
\right)}}}}{{{{{\rm{k'}}}^2} - {k^2} - i\varepsilon
}}d{\bf{k'}}d{\bf{r'}}
\end{equation}
If we define a spectral current density
\begin{equation}\label{Jk1}
{\bf{J}}\left( {\bf{k'}} \right) = \int\limits_V^{} {}
{\bf{J}}\left( {\bf{r'}} \right){e^{ - i\left( {{\bf{k'}} \cdot
{\bf{r'}}} \right)}}d{\bf{r'}},
\end{equation}
the expression (\ref{BB}) can be rewritten as follows:
\begin{equation}\label{BB1}
\textbf{B}\left( {\bf{r}} \right) =  - i\frac{{{\mu _0}}}{{8{\pi
^3}}}\int\limits_{{\bf{k'}}}^{} {} \left[ {{\bf{k'}} \times
{\bf{J}}\left( {{\bf{k'}}} \right)} \right]\frac{{{e^{i{\bf{k'}}
\cdot {\bf{r}}}}}}{{{{{\rm{k'}}}^2} - {k^2} - i\varepsilon
}}d{\bf{k'}}
\end{equation}

When deriving (\ref{BB1}), the only implicit assumption we made
was the existence of the spectral current density (\ref{Jk1}). It
can be rigorously proved that for any physical distribution of the
current density $\textbf{J}(\textbf{r})$ in a restricted volume,
the spectral density $\textbf{J}(\textbf{k})$ always exists. In
this case, $\textbf{J}(\textbf{k})$ is the integer function with a
bounded spectrum.

In a far-field region the expression (\ref{BB1}) can be
substantially simplified. In this region the electromagnetic waves
are essentially plain waves with  the only wave vector
$\textbf{k}_r$, which is directed along the vector $\textbf{r}$:
$\textbf{k}_r=\frac{2\pi}{\lambda}\frac{\textbf{r}}{r}$.
Therefore, we may take the quantity $\textbf{k}\times
\textbf{J}(\textbf{k})$ at this point out of the integral in
(\ref{BB1}) to obtain:

\begin{equation}\label{BB2}
\textbf{B}\left( {\bf{r}} \right) =  - i{\mu _0}{\left. {\left[
{{\bf{k}} \times {\bf{J}}\left( {\bf{k}} \right)} \right]}
\right|_{{\bf{k}} = {{\bf{k}}_r}}}\int\limits_{\bf{k}}^{}
{\frac{{d{\bf{k'}}}}{{{{(2\pi )}^3}}}} \frac{{{e^{i{\bf{k'}} \cdot
{\bf{r}}}}}}{{{{{\rm{k'}}}^2} - {k^2} - i\varepsilon }}\nonumber
\end{equation}
\begin{equation}
 =  - i\frac{{{\mu _0}}}{{4\pi }}{\left. {\left[ {{\bf{k}} \times
{\bf{J}}\left( {\bf{k}} \right)} \right]} \right|_{{\bf{k}} =
{{\bf{k}}_r}}}\frac{{{e^{i{k_r}r}}}}{r}
\end{equation}
where
\begin{equation}\label{kr}
{\bf{k}_r} = {\textbf{e}_x}{k}\sin \left( \theta  \right)\cos
\left( \varphi  \right) + {\textbf{e}_y}{k}\sin \left( \theta
\right)\sin \left( \varphi  \right) + {\textbf{e}_z}{k}\cos \left(
\theta \right),
\end{equation}
$\textbf{e}_x,\textbf{e}_y,\textbf{e}_z$ are unit vectors of
Cartesian coordinate system, and $k=2\pi/\lambda$.

Spherical angles $\theta$ and $\varphi$ in (\ref{kr}) coincide
with those of vector $\textbf{r}$ since in a far-field region
vector $\textbf{k}$ is directed along $\textbf{r}$.


\begin{thebibliography}{9}
\bibitem{Dress53} G. Dresselhaus, A. F. Kip, Ch. Kittel,  Observation of Cyclotron
Resonance in Germanium Crystals. Phys. Rev. \textbf{92}, 827
(1953).

\bibitem{Dress55} G. Dresselhaus, A. F. Kip , C. Kittel,  Cyclotron Resonance of
Electrons and Holes in Silicon and Germanium Crystals, Phys. Rev.
\textbf{98}, 368 (1955).

\bibitem{Latt56} J. M. Lattinger, Quantum Theory of Cyclotron Resonance in
Semiconductors: General Theory, Phys. Rev. \textbf{102}, 1030
(1956).

 \bibitem{Suzuki74} K. Suzuki, J. C. Hensel, Quantum Resonances
 in the Valence Bands of Germanium. I. Theoretical Considerations,
 Phys. Rev. \textbf{B 9}, 4184 (1974).

\bibitem{Ivanov78} V. I. Ivanov-Omskii, L. I. Korovin,  and E. M.
Shereghii, Phonon-Assisted Cyclotron Resonance in Semiconductors
phys. stat. sol. (b) \textbf{90}, 11 (1978).

\bibitem{Kano06} J. Kano and N. Miura, Cyclotron Resonance in High Magnetic Fields,
in \emph{High Magnetic Fields Science and Technology}, edited by
F. Herlach and N. Miura (World Scientific, London, 2006). Vol. 3,
p. 61.



\bibitem{Drach12} O. Drachenko and M. Helm, Cyclotron Resonance Spectroscopy,
in \emph{Semiconductor Research: Experimental Techniques}, edited
by A. Patane and N. Balkan (Springer Verlag, Berlin, 2012), p.
283.

\bibitem{Peet06} F. M. Peeters, Theory of Electron-Phonon Interactions in
Semiconductors, in \emph{High Magnetic Fields Science and
Technology}, edited by F. Herlach and N. Miura, (World Scientific,
London, 2003) Vol. 3, p. 23.

\bibitem{Horing13} \emph{Low Dimensional Semiconductor Structures: Characterization,
Modeling and Applications}, edited by H. {\"U}nl{\"u} and  N. J.
M. Horing (Springer Verlag, Berlin, 2013).

 \bibitem{Zhang14} Qi Zhang, T. Arikawa, E. Kato, J. L. Reno, Wei Pan,
 J. D. Watson, M. J. Manfra, M. A. Zudov, M. Tokman, M. Erukhimova,
 A. Belyanin, and J. Kono,  Superradiant Decay of Cyclotron Resonance
 of Two-Dimensional Electron Gases, Phys. Rev. Lett. \textbf{113}, 047601 (2014)

\bibitem{Zhang16} Qi Zhang, M. Lou, X. Li, J. L. Reno, W. Pan, J. D. Watson, M.
J. Manfra, and J. Kono, Collective, Coherent Single Photon
Superradiance, and Ultrastrong Coupling of 2D Electrons with
Terahertz Cavity Photons, Nat. Phys. \textbf{12}, 1005 (2016).

\bibitem{Laurent15} T. Laurent, Y. Todorov, A. Vasanelli, A. Delteil, C.
Sirtori, I. Sagnes, and G. Beaudoin, Superradiant Emission from a
Collective Excitation in a Semiconductor Phys. Rev. Lett.
\textbf{115}, 187402 (2015).

\bibitem{Cong16} K. Cong, Qi Zhang, Y.Wang, G. T. Noe II, A. Belyanin, and J. Kono,
Dicke Superradiance in Solids, J. Opt. Soc. Am. \textbf{B33}, C80
(2016).

\bibitem{Dicke54} R. H. Dicke,  Coherence in Spontaneous
Radiation Processes, Phys. Rev.\textbf{ 93}, 99 (1954).



\bibitem{Scully09} M. O. Scully and  A. A. Svidzinsky, The Super of
Superradiance, Science \textbf{325}, 1510 (2009).

\bibitem{Scully06}M. O. Scully, E. S. Fry, C. H. R. Ooi, and K.
W{\'o}dkiewicz, Directed Spontaneous Emission from an Extended
Ensemble of N Atoms: Timing Is Everything, Phys. Rev. Lett.
\textbf{96}, 010501 (2006).

\bibitem{Svidz08a}A. Svidzinsky and Jun-Tao Chang, Cooperative Spontaneous Emission as
a Many-Body Eigenvalue Problem, Phys. Rev. \textbf{A77}, 043833
(2008).

\bibitem{Scull09} M. O. Scully, Collective Lamb Shift in Single Photon
Dicke Superradiance, Phys. Rev. Lett. \textbf{102}, 143601 (2009).

\bibitem{Svidz08b} A. A. Svidzinsky, Jun-Tao Chang, and M. O. Scully,
Dynamical Evolution of Correlated Spontaneous Emission of a Single
Photon from a Uniformly Excited Cloud of N Atoms, Phys. Rev. Lett.
\textbf{100}, 160504 (2008).

\bibitem{Svidz09} A. A. Svidzinsky, and M. O. Scully, Evolution of
Collective N Atom States in Single Photon Superradiance: Effect of
Virtual Lamb Shift Processes, Optics Communication \textbf{282},
2894 (2009).

\bibitem{Nef17} N. E. Nefedkin, E. S. Andrianov, A. A. Zyablovsky, A. A. Pukhov,
A. P. Vinogradov, and A. A. Lisyansky, Superradiance of non-Dicke
States, Opt. Exp. \textbf{25}, 2790 (2017).

\bibitem{Klitzing80} K. v. Klitzing, G. Dorda, M. Pepper,  New Method for
High-Accuracy Determination of the Fine- Structure Constant Based
on Quantized Hall Resistance, Phys. Rev. Lett. \textbf{45}, 494
(1980).

\bibitem{Yu02} P. Y. Yu and M. Cardona,  \emph{ Fundamentals of Semiconductors.
 Physics and Materials Properties}. 3-rd ed. (Springer, New York, 2002).

\bibitem{ref}This value of magnetic field will be used for the
estimations throughout the paper.



\bibitem{Moiseev15} A. G. Moiseev, Spectrum of Holes in the Germanium and Silicon
Single Crystals in a  Quantizing Uniform Magnetic Field, Russ.
Phys. J. \textbf{ 57}, 1251 (2015).

\bibitem{Bir74} G. L. Bir, G. E. Pikus, \emph{Symmetry and Strain-induced Effects in
Semiconductors}, (Wiley, New York, 1974).



\bibitem{Auerrbach11} N. Auerbach and V. Zelevinsky, Superradiant Dynamics,
Doorways, and Resonances in Nuclei and Other Open Mesoscopic
Systems. Rep. Progr. Phys. \textbf{74}, 106301 (2011).

\bibitem{Greenberg15}
Y. S. Greenberg, A. A. Shtygashev, Non-Hermitian Hamiltonian
Approach to the Microwave Transmission Through One-dimensional
Qubit Chain, Phys, Rev. A \textbf{92}, 063835 (2015).

\bibitem{Stratton} J. A. Stratton, \emph{Electromagnetic Theory}
(Wiley,New York, 2007).



\end{thebibliography}
\end{document}